\documentclass[twoside]{LCWS11}
\usepackage[latin1]{inputenc}
\usepackage{graphics,epsfig,color}
\usepackage{epstopdf}
\usepackage{wrapfig,rotating}
\usepackage{amssymb,amsmath,array}
\usepackage{url}
\usepackage{cite}
\begin{document}
\title{
A large scale prototype for a SiW electromagnetic calorimeter
   for a future linear collider} 
\author{Roman P\"oschl$^1$ on behalf of the CALICE Collaboration
\vspace{.3cm}\\
1-  Laboratoire de l'Acc\'{e}l\'{e}rateur Lin\'{e}aire\\
Centre Scientifique d'Orsay, B\^atiment 200\\ 
Universit\'{e} de Paris-Sud XI, CNRS/IN2P3\\ 
F-91898 Orsay Cedex, France
}
\maketitle
\begin{abstract}
The CALICE collaboration is preparing large scale prototypes for highly granular calorimeters for detectors to be operated at a future linear electron positron collider. 
Currently a prototype of a SiW electromagnetic calorimeter is assembled which in terms of 
dimensions and layout meets already most of the requirements given by the linear collider 
physics program and hence the detector design. 
\end{abstract}
\section{Introduction}
The next major worldwide project in high energy particle physics will be a linear electron positron collider at the TEV scale. This machine will complement and extend the scientific results of the LHC currently operated at CERN. The most advanced proposal for this linear collider is the International Linear Collider (ILC). Here, electron and positrons will be collided at centre-of-mass energies between 0.2 and 1 TeV. The detectors, which will be installed around the interaction point are required to achieve a jet energy resolution of $30\%/\sqrt{E}$, thus a factor two better than the energy resolution achieved for a typical detector at LEP. The reconstruction of the final state of the $e^{+}e^{-}$ will be based on so-called particle flow algorithms (PFA). The goal is to reconstruct every single particle of the final state, which in turn demands a perfect association of the signals in the tracking systems with those in the calorimeters. As a consequence this requires a perfect tracking of the particle trajectories even in the calorimeters. To meet these requirements, the detectors have to cover the whole solid angle and have to feature an unprecedented high granularity.

\section{Towards a technological prototype for the SiW electromagnetic calorimeter}
The application of PFA requires a perfect reconstruction of the particle trajectories in the calorimeter. For this
the calorimeters have to be placed inside the coil of the super-conductive solenoid of the detectors. This puts tight
constraints on the available space for the installation of the detectors. The design of the detector components and notably
that of the calorimeters has to take the following guidelines into account
\begin{itemize}
\item Optimisation of the number of calorimeter cells.
\item Choice of the absorber material and the infra-structural components such as cooling, power supplies, readout
 cables and the very front end electronics.
\end{itemize}


For the electromagnetic calorimeter, which surrounds the tracking chambers these criteria has lead to the choice of tungsten with a radiation length of ${\rm X_{0}}$=3.5\,mm, a Moli\`ere of ${\rm R_{M}}$=9\,mm and an interaction length of ${\rm \lambda_{I}}$=96\,mm. In the years 2005 to 2011 the CALICE collaboration has performed beam campaigns at DESY, CERN and FNAL with a {\em physics prototype} of the electromagnetic calorimeter in order to demonstrate the principle of highly granular calorimeters and to confront the concept of particle flow with real data. The first results of the data analysis have been published in three articles ~\cite{calice1, calice2, calice3}.
The next prototype, also called {\em technological prototype}, has been conceived during the year 2008~\cite{eudet} and enters now its construction phase. It addresses, more than the physics prototype, the engineering challenges which come along with the realisation of highly granular calorimeters. The key parameters of the new prototype are

\begin{itemize}
\item Size of an individual cell of 5.5x5.5\,${\rm mm^2}$.
\item A depth of  24\,${\rm X_{0}}$.
\item Thickness of an individual layer of 3.4\,mm and 4.4\,mm according to the position within the calorimeter.
\end{itemize}

\begin{figure}[h]
\begin{minipage}[l]{0.45\columnwidth}
\centerline{\includegraphics[width=1.09\columnwidth]{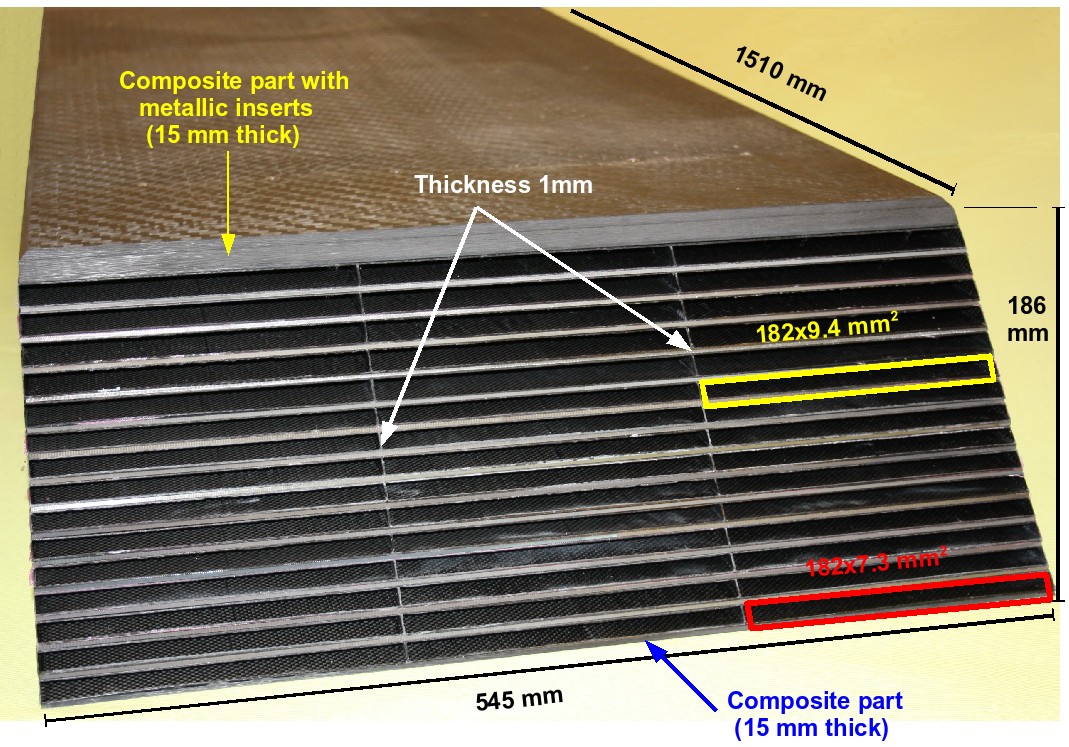}}
\caption{\sl Alveolar structure and its dimensions, which houses the sensitive parts of the CALICE SiW electromagnetic
calorimeter prototype.}
\label{fig:Alveolar}
\end{minipage}
\hfill
\begin{minipage}[l]{0.45\columnwidth}
\centerline{\includegraphics[width=1.09\columnwidth]{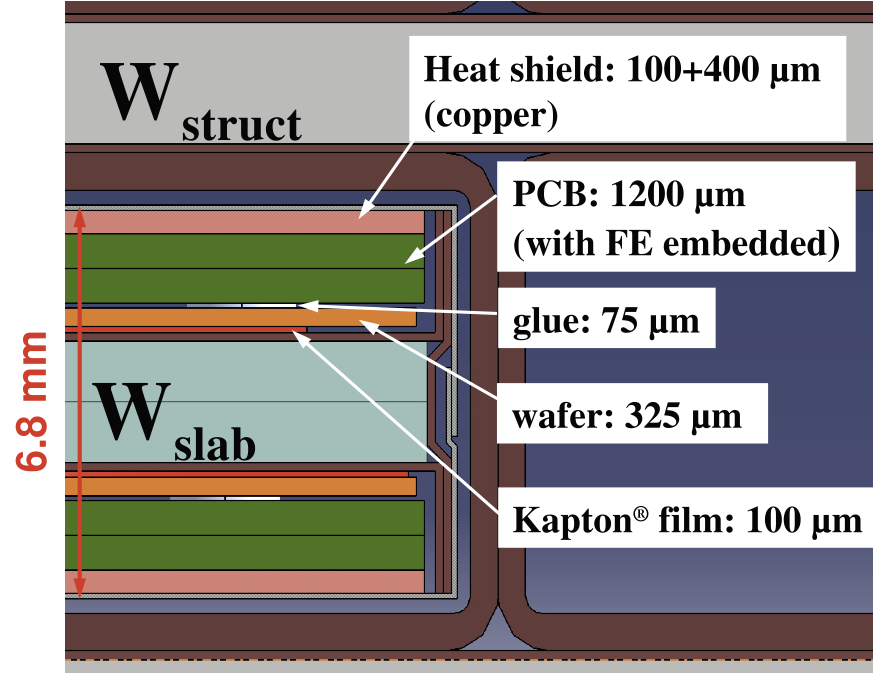}}
\caption{\sl Cross section through a slab for the CALICE SiW electromagnetic calorimeter prototype. 
The sensitive components and the very front end electronics are mounted on two sides of a tungsten carbon composite 
plate. The whole slab is embedded in alveolar layers. }
\label{fig:Slab}
\end{minipage}
\end{figure}

\begin{wrapfigure}{r}{0.5\columnwidth}
\centering
\includegraphics[width=0.45\columnwidth]{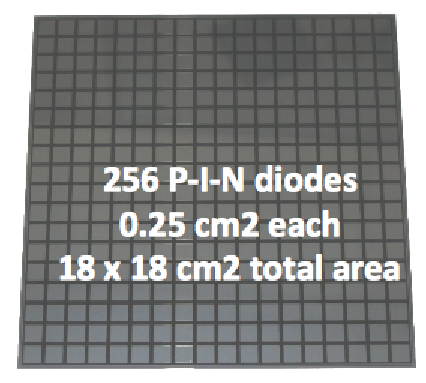}
\caption{\sl Example of a silicon wafer matrix as studied for the large scale prototype.}
\label{fig:waf}
\end{wrapfigure}

The Figures~\ref{fig:Alveolar} and~\ref{fig:Slab} show the mechanical housing, constructed during 2010, and a cross section through two calorimeter layers which form a slab. 
The mechanical housing is realised by a tungsten carbon composite, which provides at the same time the absorption medium and the mechanical rigidity of the detector. The silicon wafers are composed of high resistive silicon. An example of a silicon wafer matrix is shown in Fig.~\ref{fig:waf}. 
While in principle the manufacturing of these wafers is a well known technique, the objective is to produce these wafers at small price in order to reduce the cost since a surface as large as 3000\,${\rm m^2}$ will be needed for an ILC detector. On this contacts with major industrial partners are about to be forged. Among others the establishment of a co-operation with industry is supported within the {\em AIDA} project~\cite{aida} within the FP7 programme on research infrastructure in Europe.
The measurements with the physics prototype revealed a cross talk between the guard-ring, which surrounds the silicon wafers and neighboured silicon pads has been observed resulting in so-called {\em square events} of which the frequency increased with increasing energy of primary electrons~\cite{rnc08}. The current R\&D efforts target at the optimisation of the layout of the guard-ring. If feasible the guard-ring will even be suppressed. This would also help to meet the requirement of minimal dead space between the wafers. Additionally, a clear view will have to be established on the tolerance in the quality of the wafers. 
The wafers will be glued to the PCBs using a robotic system. Dedicated studies to realise this gluing are currently carried out. The studies comprise the pressure which can be exerted onto the wafer, the type of glue and naturally the reliability of the glue. For the latter, the experience with the physics prototype have revealed no obvious problems. It should however be noted that the constraints on the size of the glue dots were relatively relaxed. The cells of the physics prototype for example were four times larger than those of the technological prototype.  

\begin{figure}[h!]
\begin{minipage}[l]{0.45\columnwidth}
\centerline{\includegraphics[width=1.09\columnwidth]{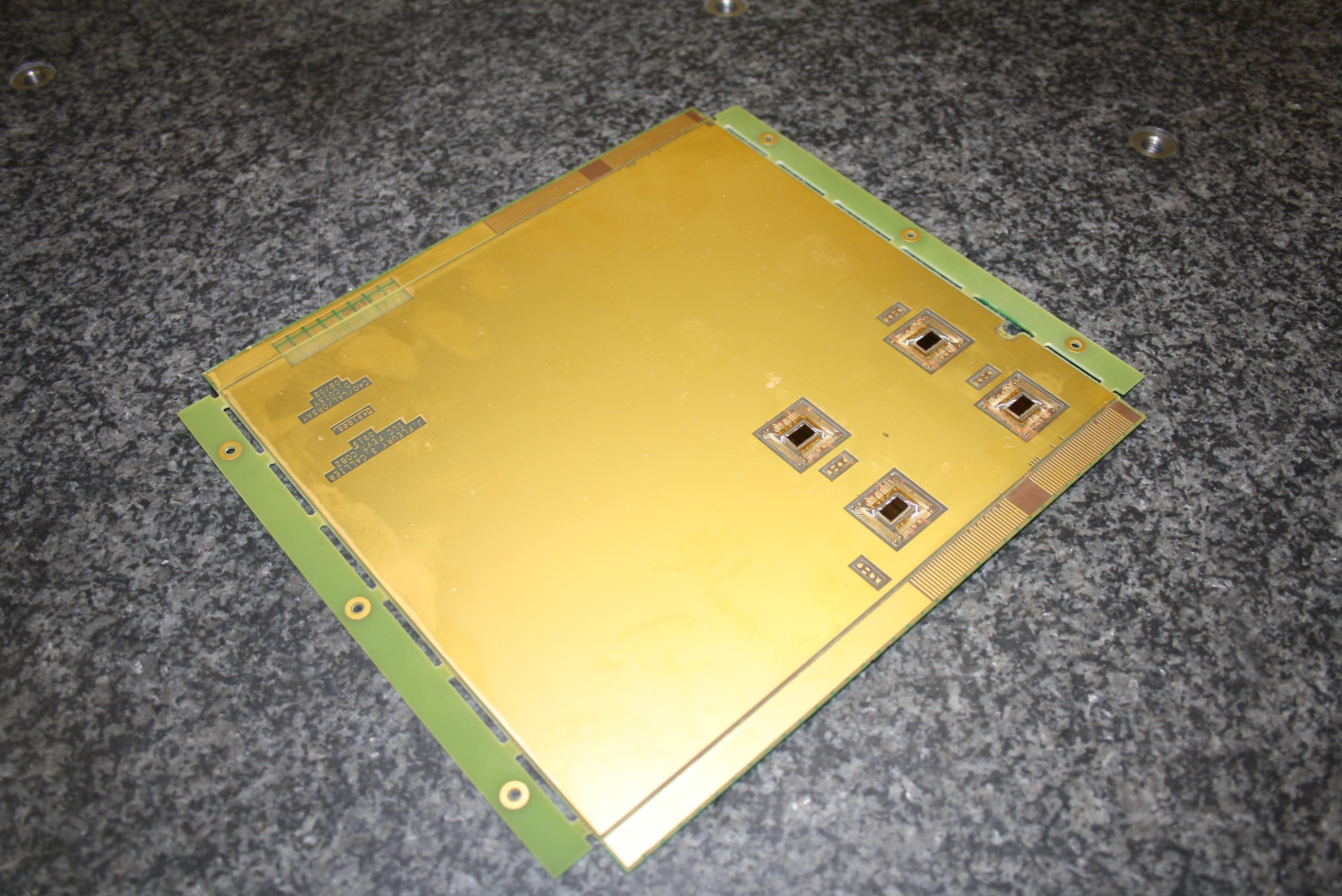}}
\caption{\sl PCB for the CALICE SiW electromagnetic calorimeter prototype with wire bonded readout ASICs.}
\label{fig:pcb}
\end{minipage}
\hfill
\begin{minipage}[l]{0.45\columnwidth}
\centerline{\includegraphics[width=1.09\columnwidth]{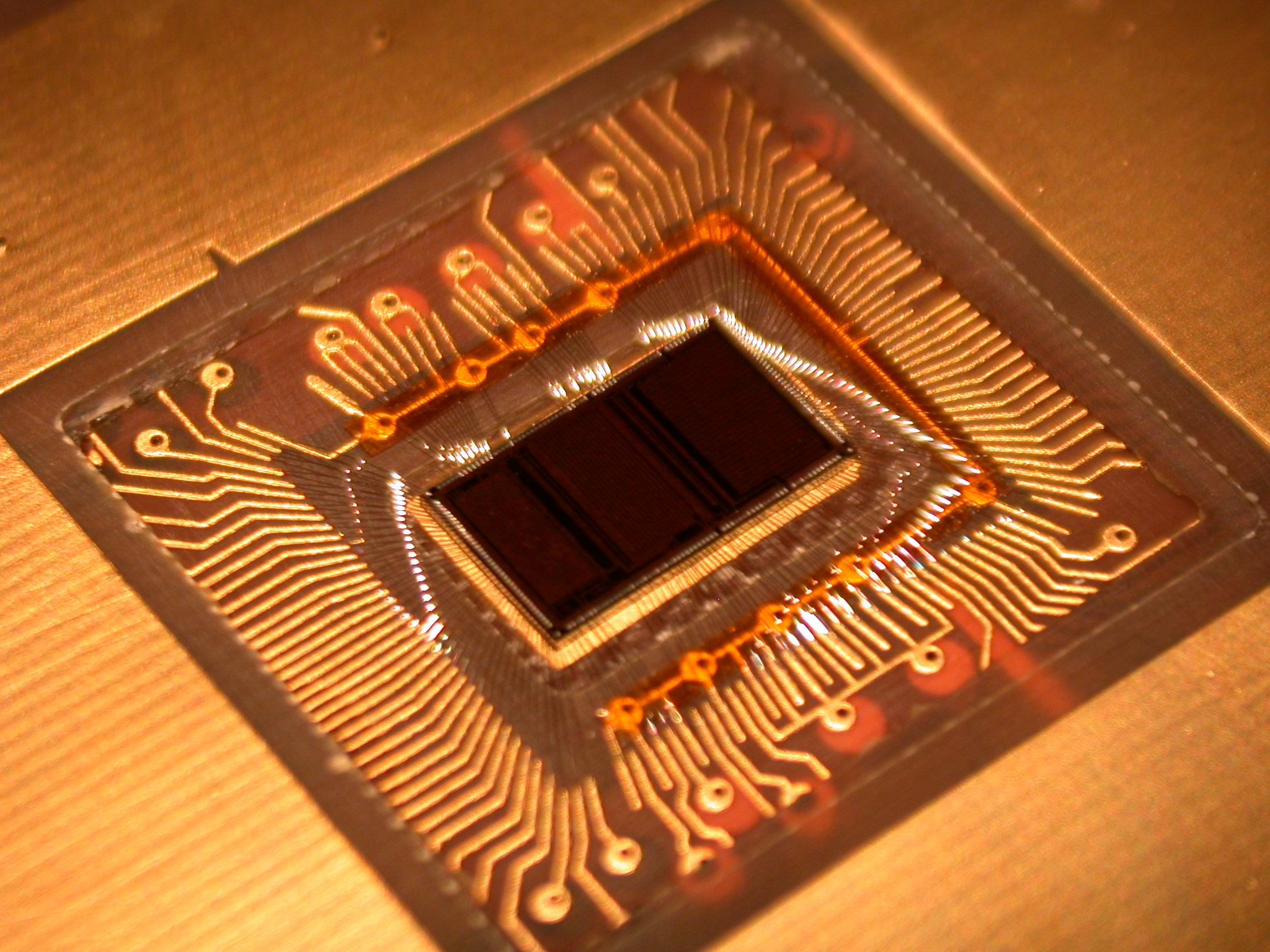}}
\caption{\sl Zoom into the zone occupied by the ASIC. }
\label{fig:asic}
\end{minipage}
\end{figure}

The final calorimeter will comprise around  $10^8$ channels in total. In order to reduce the non-equipped space in the detector, the front end electronics has to be integrated into the calorimeter layers, see Fig.~\ref{fig:Slab}, which constitutes a major challenge for the construction of the calorimeter. The space available for the readout circuits (ASICs) and the interface boards between the ASICs and the silicon wafers is about one millimetre.  This challenge is addressed in the current R\&D programme. The Figure~\ref{fig:pcb} shows a picture of four ASICs bonded onto a thin PCB. Right to that the Figure~\ref{fig:asic} shows a zoom into the area occupied by the ASICs. 
It should be noted that one of the major challenges to be solved in the near future is the planarity of the PCB. This issue is currently addressed in collaboration with industrial partners as well as by revising the entire assembly process of the detector.  Note in passing that for protection purposes the ASICs will be encapsulated. Such an encapsulation is standard in industry under the keyword GLOBTOP and has been successfully applied to circuits implemented on a prototype PCB. An alternative based on the epoxy Araldite 2020 has been developed by the research groups of CALICE.

\begin{figure}
\centering
\includegraphics[width=0.8\columnwidth]{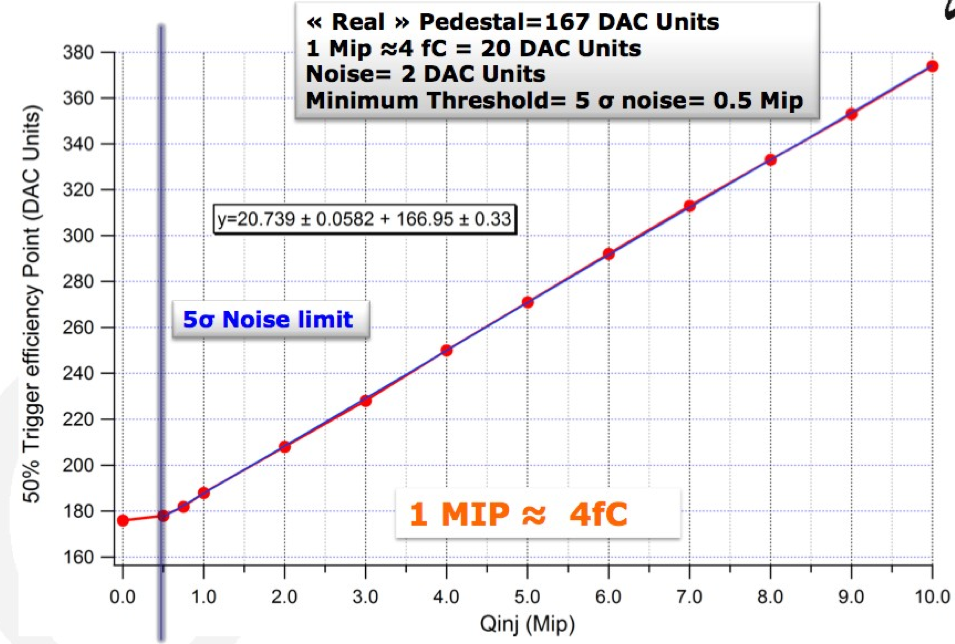}
\caption{\sl Validation of the SKIROC circuit; shown is the threshold for 50\% trigger efficiency as a function of the injected charge, in units of MIPs.}
\label{fig:skitrig}
\end{figure}

Each of the ASICs of the new prototype, dubbed SKIROC, will readout 64 calorimeter cells and realises the measuring of the analogue signal, the digitisation and the zero suppression such that only a limited number of channels are finally sent to the 
data acquisition. As an example of the excellent performance of the SKIROC circuit, the threshold of 50\% trigger efficiency as a function of the injected charge, in units of MIPs, is shown in Fig.~\ref{fig:skitrig}. In addition it is indicated that the signal over noise ratio is about eight where ten is envisaged for the circuits of the linear collider. For more details the reader is referred to~\cite{skiroc2}. A recent study reveals no undesired effect caused by embedding the circuits in the calorimeter volume as foreseen by the calorimeter design~\cite{calice5}.

\begin{figure}
\centering
\includegraphics[width=0.8\linewidth]{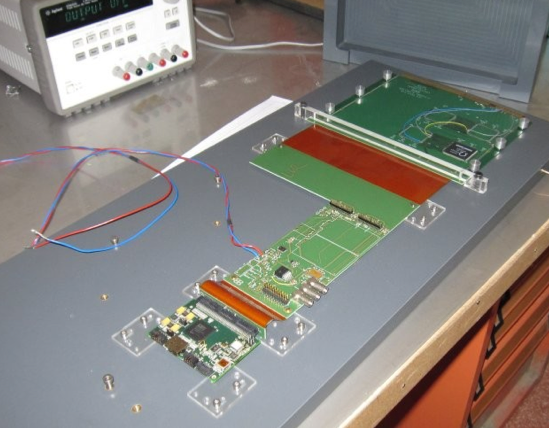}
\caption{\sl Early prototype of an ASU connected to the DAQ. The Detector Interface Card is the small board in the lower left part. Note that in contrast to Figures~\ref{fig:pcb} and~\ref{fig:asic} the ASICs are still packaged}
\label{fig:vfe}
\end{figure}

The data acquisition is based on custom made components in order to control cost and R\&D effort for a DAQ at a future linear collider. Only the part closest to the detector, called {\em Detector Interface Card}, is customised to the actual detector type. An operational set-up containing prototype versions of the front end electronics and the DAQ system is shown in Figure~\ref{fig:vfe}.

Due to the limited space available for cooling devices the heat dissipation of the ASICs has to be minimised. In 
addition to the cooling, the heat dissipation will be reduced by a novel technique called {\em Power Pulsing}. Here, the front end electronics will only be fully enabled during the millisecond of a bunch train of particles. Such a bunch train contains about 3000 particle bunches separated by around 300\,ns. During the around 200\,ms between these bunch trains, the bias currents of the front end electronics will be shut down. Clearly, this novel technique will require detailed studies in order to assure that the signal quality of each channel remains constant after each powering cycle. Power pulsing has been tested successfully in beam tests for ASICs with the a similar layout as SKIROC~\cite{imad}. 

A calorimeter layer will have a length of about 1.5\,m and will be composed of several units which carry silicon 
wafers as well as the front end electronics. Great care is taken in the development of the technique to interconnect the individual units. Apart from the reliability of the 
signal transfer along the slab the interconnection must not exert mechanical or thermal stress to the wafers which are very close to the interconnection pads. Good progress
has been made and a viable solution will be presented in the course of 2012. The sensible ensemble has to be inserted into the alveolar structure which houses the calorimeter layers. The integration cradles are under development and a first integration test with a demonstrator has been successfully conducted in October 2009. For this demonstrator
a cooling system has been developed which in an upgraded form is already available for the large scale prototype.

\section{Conclusion and outlook}
The next prototype of a highly granular SiW Ecal is taking shape. These proceedings report progress for a number of components. It should be stressed that the R\&D currently focusses on finding the technical solutions needed for an operation in a linear collider detector. Starting from 2012 the first units
of the detector will be exposed to high energy particle beams. These tests are a vital support for the ongoing R\&D. The schedule of the completion of the prototype with 30 short and one or two long ASUs is driven by the funding situation in the coming years which in turn depends also on the political situation of a future linear collider.
The solutions found for the prototype of the SiW Ecal can however also be applied in other fields of science.

\section{Acknowledgments}
This work is funded by the European Union in the 6th framework program "Structuring the European Research Area", the 7th framework program "Capacities" as well as by the CNRS/IN2P3 budget for "Quarks and Leptons", the French ANR program and the Japanese JSPS program

\section{Bibliography}



\begin{footnotesize}


\end{footnotesize}


\end{document}